# SURVEY OF UNCERTAINTY HANDLING IN CLOUD SERVICE DISCOVERY AND COMPOSITION

Nouha Khédiri[1] and Montaceur Zaghdoud[2]

[1]Faculty of Computing & IT, Northen Borders University, Rafha, KSA
nuha.khediri@nbu.edu.sa
[2]Departement of Information System, Salman b. Abdulaziz University, KSA
zaghdoud@sau.edu.sa

## ABSTRACT

*With the spread of services related to cloud environment, it is tiresome and time consuming for users to look for the appropriate service that meet with their needs. Therefore, finding a valid and reliable service is essential. However, in case a single cloud service cannot fulfil every user requirements, a composition of cloud services is needed. In addition, the need to treat uncertainty in cloud service discovery and composition induces a lot of concerns in order to minimize the risk. Risk includes some sort of either loss or damage which is possible to be received by a target (i.e., the environment, cloud providers or customers). In this paper, we will focus on the uncertainty application for cloud service discovery and composition. A set of existing approaches in literature are reviewed and categorized according to the risk modeling.*

## 1. INTRODUCTION

Cloud is a technology that represents networks and networking and computing is another technology that represents computer-related resources, applications and services. The combination of these two technologies is the concept of "Cloud computing" [1]. According to a cloud user, since the message entered a cloud and came out, so there was no need to know where it went. As soon as the web concept has been lead into the world, the user sent a Uniform Resource Locator (URL) to the internet and the request document could come back from any place. It was not necessary to know either the storage place or the owner. The utilization of cloud computing, nowadays, was compared to the electricity network from a century ago. Without installing any software or maintaining any hardware, users can use applications and conduct operations with internet access.

The essential characteristics of cloud model [2] are five as follows: on demand self-service, broad network access, resource pooling, rapid elasticity, and measured service. In addition, for the deployment models, there are three clouds that remains a unique entity, which are private cloud, community cloud and public cloud. Besides, hybrid cloud is another type of cloud models available in the market and it involves two or more clouds.
In cloud computing, there are three fundamental service models that will be described in details next section such as Infrastructure as a Service (IaaS), Software as a Service (SaaS), and Platform as a Service (PaaS).

Day after day, the number of cloud service providers increases and causes the competition in cloud market. The user can compare, evaluate and choose the cloud providers offering in the market which fulfil his requirements, but this task is a tedious and time consuming.

As cloud computing coming more popular, more and more organizations want to move to the cloud. Therefore, a risk assessment is needed to minimize the costs of implementing and maintaining controls and to help avoid surprises [3].

The terms risk and uncertainty are often used interchangeably. They are based on lack of certainty. When we talk about risk, we talk about unknown outcomes. It is the possibility of loss. Risk is quantifiable unlike uncertainty. The latter occurs when we have no idea about the outcome at the time when the decision is made [4].

On the one hand, the lack of knowledge and poor or imperfect information about all the inputs are the main causes of the uncertainty in decision making process. On the other hand, when the service provider try to describe the cloud service in a standard format, unpredictable future factors, like hacking leading to data breach or to data loss, are responsible for incomplete information, hence the need for uncertainty handling in cloud service discovery and composition. So, a good choice of cloud services conducts to a good result for the client, or a good composition between different cloud services.

In this paper, we will survey relevant contributions to the cloud service discovery and composition fields, which are inspired by the researches in cross-uncertainty. Particularly, specific interest is given to three known theories: probability theory, belief function theory and possibility theory. The remainder of this paper is organized as follows. Section 2 describes the cloud computing services. In section 3, we present the uncertainty theory and cloud risk modeling. Cloud Services Discovery and Composition are presented respectively in Section 4 and 5. Section 6 is the discussion.

## 2. CLOUD COMPUTING SERVICE

Cloud services [1] are wilfully presented within a narrow perspective of cloud applications. All the cloud services allow users to run applications and store data online. However, a different level of user control and flexibility is offered.

### 2.1. Type of Cloud Services

#### 2.1.1. Infrastructure as a Service (IaaS)

IaaS is a way of delivering cloud computing infrastructure including hardware such as servers, networking hardware and software like operating system, middleware and applications [5]. The provider of the service is the one who owns the equipment and is in charge of its management. As long as there is internet and whenever the consumer needs, the resources are available and it can be used [2]. Two types of services distinguished in [6] that allow higher level services to automate setup: "Physical Resource Set" and "Virtual Resource Set".

#### 2.1.2. Software as a Service (SaaS)

The SaaS model helps the consumer to pay only the software/service he uses. This mode eliminates the need to install and run the application on the customer's local computers. In SaaS layer [5], all the applications that give a direct service to the consumer are found. As shown in [6], the services of SaaS layer are: "Basic Application Services" and "Composite Application Services". The Basic and Composite services are categorized into Application Services, which comprise the highest level building blocks for end-user applications running in the Cloud.

#### 2.1.3. Platform as a Service (PaaS)

PaaS as the name suggests, provides computing platforms for building and running custom applications. It helps developers to speed the development of their applications and therefore save money [2]. The consumer does not manage or control the underlying cloud infrastructure including network, servers, operating systems, or storage, but has control over the deployed applications and possibly application hosting environment configurations. PaaS environment provide a complete operational environment for users to deploy and run their applications. The work in [6] categorizes the services of PaaS level into: "Programming Environments" and "Execution Environments".

Cloud computing does not come without uncertainty. When the data of company is located on the cloud, it is threatened by hackers.

## 3. UNCERTAINTY THEORY AND CLOUD RISK MODELING

For most real-world problems, uncertainty and imprecision are both often inherent in modeling knowledge. Uncertainty can result from some errors and hence from non-reliability or from different background knowledge [7]. According to [8], uncertainty is associated with the measurement errors and resolution limits of measuring instruments, at the empirical level. At the cognitive level [8], it emerges from the vagueness and ambiguities in natural language. At the social level, uncertainty is both created and maintained for various objectives by people, for example privacy and secrecy purposes [8].

### 3.1. Theories of Uncertainty

#### 3.1.1. Probability Theory

Probability theory studies the random phenomena behaviour. However, probability distribution [9] describes population probabilistic characteristics. Moreover, probability theory is a quantitative way of dealing with uncertainty which often affect several areas of life.

Given an event $A$, Probability of event $A$:

$$P(A) = \frac{Number\ of\ outcomes\ favorable\ to\ A}{Total\ number\ of\ possible\ outcomes} \qquad (1)$$

Probability theory has been developed based upon three known axioms:

$$
\begin{aligned}
&(i) \quad P(A) \geq 0 \quad (Probability\ is\ a\ nonnegative\ number) \\
&(ii) \quad P(\Omega) = 1 \quad (Probability\ of\ the\ whole\ set\ is\ unity) \\
&(iii) \quad IF\ A \cap B = \varphi,\ then\ P(A \cup B) = P(A) + P(B)
\end{aligned}
\qquad (2)
$$

#### 3.1.2. Belief Function Theory

Arthur P. Dempster is the first one who introduces the belief function theory [10] in the context of statistical inference. Then it was developed by Glenn Shafer and the theory was named after their names "Dempster-Shafer theory". The theory has another name 'Evidence theory'.

Belief function theory is based on two ideas as mentioned in [11]:

- The idea of achieving levels of belief for one issue from subjective probabilities for another related issue,
- Dempster's rule for relating such levels of belief when they depend on independent items of evidence.

In the first idea, the degree of belief, also referred to as a mass, is correspond to a belief function rather than a Bayesian probability distribution. The values of probability are assigned to sets of possibilities rather than single events.

As for the theory of Dempster-Shafer, its masses are assigned to the entire non-empty subsets of the entities that comprise a system. The framework proposed by Shafer allows propositions to be displayed as intervals which are bounded by two values. This will be explained bellow [12]:

- *belief* (denoted Bel): is constituted by the sum of the masses of all sets enclosed by an hypothesis. Belief measures the evidence strength in favour of a set of propositions. It ranges from 0 that indicate no evidence to 1 that denotes certainty.
- *plausibility* (denoted Pl): is an upper bound on the possibility of the truth of the hypothesis. It is defined to be *Pl(s)=1-Bel(~s)* with *belief ≤ plausibility*.

### 3.1.3. Possibility Theory

Because probability theory is not able to model uncertainty which result from lack or incomplete of knowledge, Zadeh [13] introduced possibility theory that differs from probability theory by the use of two sets functions respectively called possibility measure which is a maxitive set-functions and necessity measure that is a minitive set-functions [14]. In few words, for each event A from a subset S, the uncertainty may be evaluated through two levels: the possibility $\Pi(A)$, denoting to what extent the occurrence of event A if possible, and the certainty $N(A)$, denoting to what extend the occurrence of event A is certain, with:

$$\forall A \subseteq S, N(A) \leq \Pi(A) \qquad (3)$$

and

$$\forall A \subseteq S, N(A) = 1 - \Pi(\overline{A}) \qquad (4)$$

If the occurrence of *A* is certain $\Pi(A)=1$ and $N(A)=1$; if the occurrence of *A* is impossible: $\Pi(A)=0$ and $N(A)=0$. If there is no knowledge available: $\Pi(A)=1$ and $N(A)=0$ (the event is possible and not necessary).

Dubois and Prade in [15] confirmed that possibility theory is a rough non-numerical probability version. For them, it is a simple methodology to reasoning since its probabilities are not precise.

### 3.2. Cloud Risk Modeling

When something happens and negatively affects the outcome, it is called risk [3]. Risks related to the use of cloud are cumulative according to the type of service to which they relate: IaaS presents a series of risks which are specific and are related to the management infrastructure; PaaS show the same risks with, in addition, those that are specific to this type and which are related to the management system software (i.e., it limits developers to provider languages and tools); Finally, the SaaS model has, in moreover, the specific risks related to data management and applications which not always suitable for business use.

The risk management has three general steps mentioned in [16]. First, the risk data is identified before it becomes problems. Then, it is analysed and converted to decision-making information. Finally, the status of risk is controlled and some actions are taken into consideration to reduce the risk.

As many approaches based on probabilistic methods have been proposed to tackle the uncertainty in practical risk assessment and this is due by the Bayesian formulation for the treatment of rare events and poorly known processes. However, a number of theoretical and practical challenges seem to be still somewhat open [17]. This has sparked the emergence of a number of alternative approaches, which are:

- Use of interval probabilities due to the combination of probability analysis and interval analysis,
- Imprecise probability is considered to provide a more complicated representation of uncertainty,
- Random sets in the two forms proposed by Dempster and Shafer,
- Possibility theory that is formally a special case of the imprecise probability and random set theories.

## 4. CLOUD SERVICE DISCOVERY

With the growing number of various web services over the internet, the efficiency and the scalability of service discovery become a critical issue for cloud computing enabled applications. To discover a cloud service provider which is reliable is very vulnerable. A system is less trusted, if it provides us with information which is insufficient in relation to its abilities. By trust it is meant faith or confidence that a promised thing will for sure be realised [18,19]. Today, it is fundamental to have confidence between the service provider and the user [20]. It is not possible for the user to check whether the service is trustful and whether there is a risk linked to insider attacks. In cloud environment, service discovery is challenging among the available heterogeneous services and the large number of service providers. Moreover, security is an enormous challenge while searching for a suitable cloud service.

Raggad [21] proposes a decision support mechanism that applies Dempster and Shafer theory [22] to devise a risk driven cloud computing project solution. The decision support model consists of the following steps; the technical team collects evidence on available and relevant cloud computing providers and brokers. This belief structure consists of a basic probability assignment for every service, on the frame of discernment of cloud providers and brokers. After given the cloud computing power and the cloud brokerage power, the technical team contact relevant auditors to collect evidence on the capabilities of the defined cloud computing power (i.e., selected cloud providers) to establish all company's IT service requirements when executing the assigned IT services. This belief structure [21] is used to make sure that the cloud computing project solution adopted above can in fact establish all mandated IT service requirements. The last step is computing the total expected risk of acquiring cloud power based on which a decision can be made whether to acquire cloud computing or process candidate IT services internally.

Rajesh et al. [23] developed a new approach that uses the belief net to cleverly discover the appropriate and relevant web service relying on the available QoS parameters. They use ontology, as a semantic search engine, which helps the cloud customer to discover the service that would fulfil a specific need. This ontology offers more precise and relevant results by using the semantic annotation that helps to bridge the ambiguity of the natural language.

In [24] a probabilistic flooding based method is proposed, which is easily to implement and not sensitive to the dynamic change in the network. This method combine Simple Additive Weighting (SAW) technique and skyline filtering, that is proposed to perform QoS-aware service discovery over a service registering-enabled P2P network, according to the user's functional requirements and QoS constraints. The Simple Additive Weighting is a technique that could be used to select a QoS-optimal service among a set of functionality equivalent services, taking into consideration of QoS properties of these services. There are still some limitations. Firstly, the time cost of service discovery for different service query under the same traffic situation, is not determined in a quantitative way. Moreover, simulation experiment is the only way to illustrate the performance of the method of service discovery. Finally, more uncertain factors could affect the performance of this method, such as node crush and traffic congestion.

In a novice research [20] trust mechanism is integrated into cloud service discovery process. As a solution for the specific risk problems, trust is used. Trust means the faith and the confidence that the service will surely be supplied as convenient. The Bayesian network was adopted for formulating this model. The probabilistic dependencies among the variables in the cloud service discovery field were captured as conditional probability [20]. Hence the model was illustrated as a series of connected Bayesian Network that is illustrated as an acyclic directed graph.

Relying on the state of the cloud infrastructure, cloud provider might not be the most reliable at a well precise moment. The limitation of the studies of efficiency of how to exploit a single cloud system let to the deployment of service composition phase.

## 5. CLOUD SERVICE COMPOSITION

Composition [25] is called static when it is done manually in customer's requirement design step and the cloud service is automatically selected by the system. Nonetheless, in dynamic composition, the service selection and the process model are automatically done by programs. Service composition can scale Cloud computing in two dimensions: horizontal and vertical [26]. Horizontal service composition stands for the composition of services which are heterogeneous and which may be in several Clouds. Vertical service composition refers to the services which are homogenous in order to increase a given Cloud node capacity.

The failure of the composite service is engendered by the failure of individual cloud service. Composite service general reliability is the product of constituent cloud services reliability. Therefore, one unreliable cloud service can decrease the overall reliability to a very low level.

Many services are involved in an application and that can be available as a novice cloud service [27]. Any provider of this service will face a hard situation since it is essential to guarantee a particular level of QoS to end users. Simultaneously, the quality of the supplied service relies on agreements between both the partners and quality of services [28]. A composite service will be obliged to pay fines to its customers because it cannot meet the entire required request on time. In general, it is possible to speak about risk which quality of a composite web service can be affected because of problems with related services. In case the risk is important, it is a must to mitigate it.

In this part, we discuss several methods for service composition in the cloud using uncertainty. In [29] the approach uses Bayesian networks as well as probability mixtures to model service composition in order to access separately each quality of service. The consumer can then choose services according to its preferences from the varied qualities. The model illustrates the dynamism by developing the Bayesian network, which therefore positively affects the quality of the service in terms of trustfulness. Trust systems enable parties to determine the trustworthiness of participating parties [30]. The Bayesian approach illustrates the relationship of supplying a good quality of service among the composite and the constituent services. Moreover, it constantly updates trust to meet newly quality. Bayesian approach can first model the relations of the service composition. Second, distinguish between any good or bad services in any partially noticeable setting. Third, deduct conditional probabilities from the relationships. However, the Bayesian approach didn't succeed to compute the constituent services and unconditional trustworthiness. Moreover, the Bayesian approach needs at least partial observability. In its model trust, it uses only the probability unlike modern approaches which use both probability and certainty.

Reference [31] focuses on Quality of Services. QoS satisfied prediction model which is relied on a hidden Markov model (HMM). HMM is a good technique to solve problems of prediction.

[32] presents algorithms as well as a framework that simplify cloud service composition for non talented users. The authors use a combination of fuzzy logic and algorithms for composition optimization to facilitate the user's task. Moreover a ranking system for cloud service composition is provided to help users express their preferences in convenient way by using high-level linguistic rules. The main limitation in cloud service composition is the ability to find the appropriate combination which reduces the deployment cost and time, and increases the reliability.

# 6. DISCUSSION

The real internet is more dynamic, and more uncertain factors could affect the performance of the method such as traffic congestion and the qualities offered by a service instance might vary over time, sometimes rapidly. But, the important hurdles to users adopting cloud services involve security, availability, and reliability.

The study allowed us to identify the main technical of cloud service discovery and composition under uncertainty that is an important nature of trust.

As a summary for detailed analysis of the related work (see Table 1 and 2), we have identified a set of characteristics as risk modeling namely: are:

Type of Service (TS): denotes the type of cloud services which are IaaS, PaaS and SaaS.
Type of uncertainty (TU): determines which theory or technique was used to model uncertainty.
Response Time (RT): refers to the interval between the sent request and the obtained reply.
Robustness of process (RP): it deals with error management during discovery.

Table 1. Summary of Cloud Service Discovery.

| Contri-bution | Mechanism | Risk Modeling | | | |
|---|---|---|---|---|---|
| | | *TS* | *TU* | *RT* | *RP* |
| Raggad [21] | Belief Function | IT service | Dempster & Shafer Theory | – | Expected Losses |
| Lin et al. [22] | Service Registering | SaaS PaaS IaaS | Probability Theory | Small time to live | Node crush Traffic congestion |
| Rajesh et al. [23] | Bayesian Network | SaaS | Belief Network | Dynamic and Quick | Error handling is little hard |
| Akinwunmi et al. [20] | Trust mechanism | – | Probability Theory | Good | Less general risk Users' anxiety reduced |

We can address the issues in the following tables 1 and 2 for the approaches that are discussed in the previous sections.

Literature in [21], [23] and [24] points out that there is only a few approaches in cloud service discovery which are under uncertainty. In fact, it is not possible for the user to be certain about the quality of the service and its trustworthiness. Also, in [24] a small Time To Live could avoid the traffic congestion, but this reduced the number of service information replicas, which may impair the service discovery efficiency. Moreover, we note that, the existing approaches use only the belief function or the probabilistic measures but not the possibilistic ones that have been successfully applied in decision making problems in conditions of uncertainty.

There are new challenges raised by service composition in cloud [32] which are caused by the diversity of users across different geographical locations with all their different legal constraints. An evaluation of techniques for service composition in the cloud is presented below in Table II.

Table 2. Summary of Cloud Service Composition.

| Contri-bution | Mechanism | Risk Modeling | | | |
|---|---|---|---|---|---|
| | | TS | TU | RT | RP |
| Hang et al. [29] | Bayesien Network | – | Probability Theory | – | Unconditional trustworthiness |
| Wu et al. [31] | Hidden Markov | SaaS PaaS IaaS | Prediction model | Small time to live | Expected larger error |
| dastjerdi et al. [32] | Defuzzification strategy [33] | IaaS | Fuzzy Preferences | Optimized | – |

As shown in the above analysis, neither of the existing methods of cloud service composition is based on the possibility theory, which is very powerful to represent partial or incomplete knowledge [34].

These approaches have been proposed to tackle the problem of cloud service discovery and composition, but they were inadequate enough to handle the uncertainty associated with the cloud environment.

Possibility theory is distinct from probability theory, the theory of random sets, and Dempster-Shafer theory of belief and plausibility. Viewed in this prospect, the distinctness of possibility theory implies that it is in a complementary relation to probability theory and not an alternative to it. This involves that it addresses a class of issues in the management of uncertainty which are not addressed by probability theory [28].

## 7. CONCLUSIONS

Despite the cloud computing is the preeminent on-demand service system along with a "Pay-as-you-go", the service provided does not respond to our needs with a total way (100%). In other words, the service provided is not in conformity with the request. This problem brings us to the uncertainty reasoning which is important to ensure the user's satisfaction.

The aim of this research is to give a general view of any new development in cloud service discovery and composition under uncertainty. The main objective is to emphasise the interest in the risk analysis by uncertainty handling, because, both risk and uncertainty research can help to make appropriate decisions about the necessary actions in case of shortage of knowledge about the system state.

Besides for researchers, this systematic review might have implications for practitioners. They can use this review as a source in searching for relevant approaches for Cloud Service Discovery and composition under uncertainty.